\documentclass[twocolumn,10pt]{tsfp}
\usepackage{flushend}
\usepackage{graphicx}
\usepackage[authoryear,round]{natbib}
\usepackage{fancyhdr}
\usepackage{xcolor}
\usepackage[nointegrals]{wasysym}
\usepackage{MnSymbol}

\title{Characteristics of active and inactive motions in high-Reynolds-number turbulent boundary layers}

\author{Rahul Deshpande$^{1}$, Ricardo Vinuesa$^{2}$, Ivan Marusic$^{1}$
    \affiliation{$^{1}$Department of Mechanical Engineering, The University of Melbourne, Victoria 3010, Australia}
    \affiliation{$^{2}$FLOW, Engineering Mechanics, KTH Royal Institute of Technology, Stockholm, 10044, Sweden}\\	
}

\begin{document}

\maketitle   %Print title matter
\thispagestyle{fancy}

% Set the font to 9pt.
\fontsize{9}{11}\selectfont

%%%%%%%%%%%%%%%%%%%%%%%%%%%%%%%%%%%%%%%%%%%%%%%%%%%%%%%%%%%%%%%%%%%%%%
\section*{ABSTRACT}
%Townsend's attached eddies are a significant contributor to the total drag produced by a high-Reynolds-number ($Re_{\tau}$) turbulent boundary layer (TBL). The present study investigates the underlying mechanism behind this by analyzing the active and inactive components of the attached eddies postulated by \citet{townsend1976}. A recently proposed energy-decomposition scheme is implemented on TBL datasets spanning a large $Re_{\tau}$ range ($\sim$ $\mathcal{O}$(10$^{3}$)--$\mathcal{O}$(10$^{6}$)), to present empirical evidence of the role played by these two components in drag generation for the first time. It is found that while the active motions solely produce the Reynolds shear stresses, the inactive motions transport the streamwise momentum from the log region to the wall, confirming prior hypotheses in the literature.
%The significance of Townsend's attached eddies, as a substantial contributor to the total drag offered by a high Reynolds number ($Re_{\tau}$) wall flow, is well accepted in the literature.

Wall-scaled (attached) eddies play a significant role in the overall drag experienced in high-Reynolds-number turbulent boundary layers (TBLs). This study aims to delve into the underlying mechanisms driving this phenomenon by dissecting the active and inactive components of these attached eddies, as initially proposed by Townsend (1976). Employing a recently introduced energy-decomposition scheme, we analyze TBL datasets covering a wide range of Reynolds numbers ($Re_{\tau}$ $\sim$ $\mathcal{O}$(10$^{3}$)--$\mathcal{O}$(10$^{6}$)). This analysis provides empirical evidence of the distinct contributions of these components to drag generation, and reveals that while active motions are responsible solely for generating Reynolds shear stresses, inactive motions are crucial for transporting streamwise momentum from the logarithmic region to the wall, thus corroborating earlier hypotheses in the literature.

%%%%%%%%%%%%%%%%%%%%%%%%%%%%%%%%%%%%%%%%%%%%%%%%%%%%%%%%%%%%%%%%%%%%%%
\section*{INTRODUCTION \& MOTIVATION}

Flow over an airplane wing or over a ship hull corresponds to a turbulent boundary layer (TBL) at a very high friction Reynolds number ($Re_{\tau}$ $>$ 10$^{5}$; where $Re_{\tau}$ = ${U_{\tau}}{\delta}/{\nu}$ with $U_{\tau}$, $\delta$ and $\nu$ being the mean friction velocity, boundary layer thickness and kinematic viscosity, respectively), resulting in significant power requirements to overcome the skin-friction drag produced by the overlying TBL.
%Such high $Re_{\tau}$ means these vehicles require powerful engines to overcome the huge skin-friction drag/resistance offered by the TBL.
Despite such economic incentives on offer, progress in understanding the flow physics responsible for this drag has been slow owing to differing governing mechanisms in low ($Re_{\tau}$ $\lesssim$ 10$^{3}$) and high ($Re_{\tau}$ $\gtrsim$ 10$^{4}$) $Re_{\tau}$ flows, with only the former easily realizable in a laboratory \citep{marusic2021}.
While the viscous-scaled coherent motions in the near-wall region are responsible for generating drag at low $Re_{\tau}$ \citep{kim2011,corke2018}, the energy containing self-similar motions in the log/inertial region have been shown to generate a significant portion of the total drag at $Re_{\tau}$ relevant to the transportation industry and several other engineering applications \citep{deck2014,giovanetti2016,marusic2021}.
The log region is also responsible for the bulk production of the turbulent kinetic energy at these high $Re_{\tau}$ \citep{marusic2010high,smits2011}, encouraging the modelling of this region and making it an area of active research.

\begin{table*}[t]
\centering
\caption{Details of various published multi-point datasets of canonical TBLs used in the present study. LES indicates Large Eddy Simulations, HRNBLWT indicates High Reynolds Number Boundary Layer Wind Tunnel Facility at the University of Melbourne, and SLTEST refers to the Surface Layer Turbulence and Environmental Science Test (SLTEST) facility in western Utah.
Bold and underlined numbers respectively refer to the lower and upper bounds of the log region, \emph{i.e.} $z^+$ $\sim$ 2.6$\sqrt{Re_{\tau}}$ and 0.15$Re_{\tau}$.}
\label{tab1}
\begin{tabular}{c c c c c c c c c c c}
& & \\ % put some space after the caption
\hline
\hline
$Re_{\tau}$ &  & Facility/ & & Wall ($z$ = 0) & &	log region & & log region & & Reference\\
 & & Simulations & & sensor & & sensor & & range & & \\
\hline
%Custom probe\footnotemark & $uw$ & 7 & 4 \\
2000 & & LES & & - & & - & & {\bf 100} $\lesssim$ $z^+$ $\lesssim$ \underline{250} & & \citet{eitel2014}\\
15000 & & HRNBLWT & & hot-film & & cross-wire & & {\bf 300} $\lesssim$ $z^+$ $\lesssim$ \underline{2250} & & \citet{talluru2014}\\
$\mathcal{O}$(10$^{6}$) & & SLTEST & & shear sensor & & sonics & & 3360 $\lesssim$ $z^+$ $\lesssim$ 41000 & & \citet{marusic2007}\\
\hline
\hline
\end{tabular}
\end{table*}
%\footnotetext{Further details may be found in \citet{baidya2012}}

Of the several models proposed in the literature, the conceptual model by \citet{townsend1976} known as the attached eddy model (AEM), which is based on his attached eddy hypothesis, is one of the most prominent \citep{marusic2019}.
The AEM essentially models the kinematics in the log region of a canonical TBL flow by a hierarchy of inertia-dominated, geometrically self-similar eddying motions that extend to the wall and are randomly distributed in the flow field.
%The `attached' term here refers to any coherent motion that has a geometric extent scaling with its distance from the wall, $z$.
These eddies have a population density inversely proportional to their height ($\mathcal{H}$), that varies between $\mathcal{O}$($z_{l}$) $\lesssim$ $\mathcal{H}$ $\lesssim$ $\mathcal{O}$($\delta$).
Here, $z_{l}$ corresponds to the lower bound of the log region ($z^{+}_{l}$ $\approx$ 2.6$\sqrt{Re_{\tau}}$) while its viscous-scaled upper bound is generally accepted to be 0.15$Re_{\tau}$.
As per the hypothesis, such a distribution yields a logarithmic variation in the variances of the viscous-scaled wall-parallel velocity fluctuations (${\overline{u^2}}^{+}$, ${\overline{v^2}}^{+}$), while the wall-normal velocity variance (${\overline{w^2}}^{+}$) and the Reynolds shear stresses (${\overline{uw}}^{+}$) tend to be a constant with the wall-normal coordinate $z$ following:
\begin{equation}
\label{eq1}
\begin{aligned}
&{{\overline{u^2}}^+} = {B_1} - {A_1}\; \ln{\bigg(}\frac{z}{\delta}{\bigg)}, \hspace{5mm}
{{\overline{v^2}}^+} = {B_2} - {A_2}\; \ln{\bigg(}\frac{z}{\delta}{\bigg)}, \\
&{{\overline{w^2}}^+}= {B_3}\; \hspace{5mm} \text{and} \; \hspace{5mm} {{\overline{uw}}^+} = -1,
\end{aligned}
\end{equation}
with $B_{1-3}$ and $A_{1-2}$ being constants.
One can find substantial empirical evidence in support of these expressions in the literature  \citep{jimenez2008,hultmark2012,mklee2015,pirozzoli2021,deshpande2021}. %, suggesting an ever increasing credibility of the model.
The contrasting nature of variation between (${\overline{u^2}}^{+}$, ${\overline{v^2}}^{+}$) and (${\overline{w^2}}^{+}$, ${\overline{uw}}^{+}$) in (\ref{eq1}) was explained by \citet{townsend1976} via decomposition of the total flow at any $z$, in the log region, into `active' and `inactive' contributions.
%The roots for this difference lie in the spatially localized signature of the $w$-velocity from the attached eddies, in comparison to that noted for the wall-parallel velocity components.
Here, the active motions correspond to the `localized' attached eddies of height, $\mathcal{H}$ $\sim$ $\mathcal{O}$($z$), and they add to ${{\overline{u^2}}}(z)$, ${{\overline{v^2}}}(z)$, ${{\overline{w^2}}}(z)$ and ${{\overline{uw}}}(z)$.
While, the inactive contributions come from the relatively large and tall attached eddies of height, $\mathcal{O}$($z$) $\ll$ $\mathcal{H}$ $\lesssim$ $\mathcal{O}$($\delta$), that only add to ${{\overline{u^2}}}(z)$ and ${{\overline{v^2}}}(z)$ but not to ${{\overline{w^2}}}(z)$ and ${{\overline{uw}}}(z)$.
Based on the above description, the active motions can be described as the sole Reynolds shear stress producing motions in the log region \citep{deshpande2021,deshpande2021uw}.
Interested readers can refer to these papers for further and elaborate discussions on this topic.
These discussions inspire the decomposition of the attached eddy velocity fields per the following \citep{panton2007}: 
\begin{equation}
\label{eq100}
\begin{aligned}
u &= {{u}_{\rm a}} + {{u}_{\rm ia}},\\
v &= {{v}_{\rm a}} + {{v}_{\rm ia}},\\
w &= {{w}_{\rm a}},
\end{aligned}
\end{equation}
%$u$ $=$ ${u}_{\rm a}$ $+$ ${u}_{\rm ia}$, $v$ $=$ ${v}_{\rm a}$ $+$ ${v}_{\rm ia}$ and $w$ $=$ ${w}_{\rm a}$, 
where subscripts `$\rm a$' and `$\rm ia$' represent active and inactive contributions, respectively.

Given that the attached eddies are statistically significant in generating skin-friction drag in a high-$Re_{\tau}$ TBL \citep{deck2014,giovanetti2016,marusic2021}, it is worth considering how their active and inactive components individually contribute to this dynamical process.
\citet{townsend1976} described inactive motions as \emph{non-local} `swirling motions' whose ``\emph{effect on that
part of the layer between the point of observation and the wall is one of slow random
variation of `mean velocity’ which cause corresponding variation of wall stress.}''
In other words, owing to the relatively large spatial coherence of the inactive motions compared to the active motions at any $z$, \citet{townsend1976} described the former to be influencing the velocity field at all wall heights below $z$, including the wall-shear stress fluctuations (via low-frequency fluctuations).
This characteristic of the inactive motions was hypothesized by \citet{giovanetti2016} as the pathway through which the attached eddies contribute to the turbulent skin-friction drag.
They argued that the attached eddies generate Reynolds shear stress through their active part but transport the streamwise momentum to the wall through their inactive part.
This hypothesis has important implications for the future flow control schemes, but has not been tested yet due to the absence of a reliable methodology that can decompose the log region flow into its active and inactive components.
To this end, the present study demonstrates the efficacy of a recently proposed decomposition methodology \citep{deshpande2021,deshpande2021uw} on TBL datasets spanning a large-$Re_{\tau}$ range ($Re_{\tau}$ $\sim$ 10$^{3}$ -- 10$^{6}$), thereby permitting testing of the hypothesis proposed by \citet{giovanetti2016} for the first time. % at relevant/high $Re_{\tau}$.

%%%%%%%%%%%%%%%%%%%%%%%%%%%%%%%%%%%%%%%%%%%%%%%%%%%%%%%%%%%%%%%%%%%%%%
\section*{DATASETS AND METHODOLOGY}

The present study considers three published, multi-point datasets from canonical TBLs (table \ref{tab1}).
The data set at lowest $Re_{\tau}$ ($\sim$ 2000) is from the well-resolved large-eddy simulation (LES) of \citet{eitel2014}.
The intermediate $Re_{\tau}$ ($\sim$ 15000) data was acquired in the large Melbourne wind tunnel (HRNBLWT; \citep{talluru2014}) using hot-film sensors (on the wall) and cross-wire sensors. 
While the data set at highest $Re_{\tau}$ ($\sim$ $\mathcal{O}$($10^6$)) was acquired at the atmospheric surface layer facility (SLTEST) in neutrally-buoyant conditions, using a combination of sonic anemometers and a unique wall-shear stress sensor \citep{marusic2007}.
%Table \ref{tab1} lists the basic details along with the references to which these datasets correspond, which can be directly referred for the experimental details.
Hence, each of these datasets provide access to the time-series of the instantaneous wall-shear-stress fluctuations (${\tau}_{w}$) acquired/computed synchronously with the \{$u$,$w$\} fluctuations in the log region, by a probe placed vertically above the wall-sensor.

Notable differences between these data sets, however, are in the length/duration of the available time series and the multi-point nature of data acquisition.
In case of the HRNBLWT data, ${\tau}_w$ was acquired simultaneously with a cross-wire probe, which was traversed across the canonical TBL to measure long time series of $u,w$ at logarithmically spaced $z$-locations. 
As a consequence, the $u,w$-fluctuations acquired at each $z$-location are statistically independent from one another.
Further, the time series data from the HRNBLWT was acquired for a sufficiently long sampling time ($t_{\rm samp}$ $\sim$ 360\:s), such that ${t_{ \rm samp}}{U_{\infty}}/{\delta}$ $\gtrsim$ 30000, which ensures converged large-scale spectral features.
In contrast, the duration of the available time series from the LES and SLTEST data set are respectively limited to ${t_{\rm samp}}{U_{\infty}}/{\delta}$ $\approx$ 243 and 175, owing to computational/experimental limitations.
But it is worth highlighting that both these data sets are unique in the sense that ${\tau}_w$ as well as $u,w$-fluctuations (across the log region) are acquired synchronously.
Consequently, in this study, we will first demonstrate consistency in the trends exhibited by the spectral estimates from all the three data sets, to confirm that differences in statistical convergence do not influence these statistics qualitatively.
These spectral estimates will then be used to estimate the active and inactive components of the velocity fluctuations from the HRNBLWT data.
Finally, we will analyze the synchronously acquired $u,w$-fluctuations from the LES and SLTEST data to correlate the instantaneous features of their active and inactive components with ${\tau}_w$-fluctuations.

	\begin{figure*}
	\centering
	\includegraphics[width=1.05\textwidth]{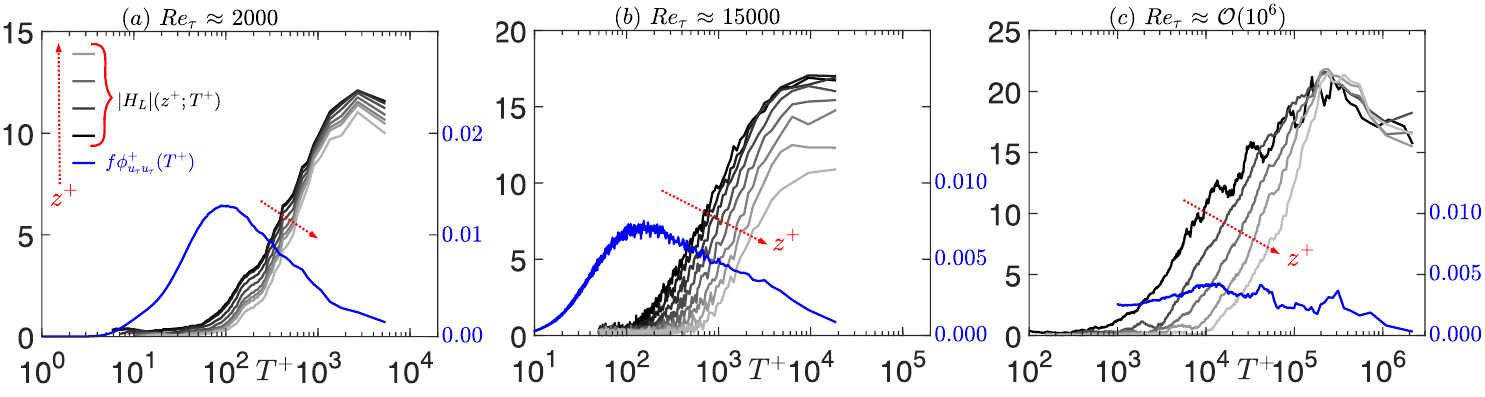}
	\caption{
Modulus of the linear transfer kernel ($|H_L|$($z^+$;$T^+$); equation ) estimated for the $Re_{\tau}$ $\approx$ (a) 2000, (b) 15000 and (c) $\mathcal{O}$(10$^{6}$) data sets.
$|H_L|$ is depicted only for $z^+$ corresponding to the log region in the wall-normal range indicated in table \ref{tab1}, and its ordinate is on the primary vertical axis (left).
The solid blue line represents the premultiplied spectra of the friction velocity ($f{{\phi}^+_{{u_{\tau}}{u_{\tau}}}}$), ordinate for which is on the secondary vertical axis (right).}
	\label{fig1}
	\end{figure*}

In this study, we implement the spectral linear stochastic estimation (SLSE)-based decomposition methodology proposed by \citet{deshpande2021} and \citet{deshpande2021uw} to obtain the inactive component ($u_{\rm ia}$) at $z^{+}$ in the log region:
\begin{equation}
\label{eq13}
%\begin{split}
{{\widetilde{u}}_{\rm ia}}(z^{+};{f^{+}}) = {{H_{L}}(z^{+};{f^{+}})}{{\widetilde{u}}_{\tau}}({f^{+}}), 
%\end{split}
\end{equation}
where ${{\widetilde{u}}_{\tau}}({f^{+}})$ = $\mathcal{F}$(${{u}^{+}_{\tau}}$($t^{+}$)) is the Fourier transform of the skin-friction-velocity fluctuation, ${u_{\tau}}$ in time 
(where, ${u_{\tau}}$ = $\sqrt{{{\tau}_{w}}/{\rho}}$).
The time domain equivalent of $\widetilde{u_{\rm ia}}$ is estimated simply by the inverse Fourier transform.
Note that $H_{L}$ in (\ref{eq13}) corresponds to a complex-valued, scale-specific transfer kernel between the synchronously-acquired $\widetilde{u}$($z$) and $\widetilde{u_{\tau}}$, and is defined as:
\begin{equation}
\label{eq14}
{H_{L}}(z^{+};{T^{+}}) = \frac{ \{ \widetilde{u}(z^{+};{T^{+}}){{\widetilde{u_{\tau}}}^{\ast}}({T^{+}}) \} }{ \{ {{\widetilde{u_{\tau}}}({T^{+}})}{{\widetilde{u_{\tau}}}^{\ast}}({T^{+}}) \} }.
\end{equation}
Considering $|H_{L}|$ is a statistically averaged quantity, it can estimated at various $z^+$ in the log region for all the three multi-point data sets listed in table \ref{tab1}. 
Figure \ref{fig1} depicts the absolute value of $H_{L}$ at various $z^+$ (\emph{i.e.}, $|H_{L}|$($z^+$)) and compares it with the respective premultiplied spectra of $U_{\tau}$.
Essentially, this figure presents a visual description of both the parameters required to estimate 
${{\widetilde{u}}_{\rm ia}}$ in equation (\ref{eq13}).

The consistent variation of $|H_{L}|$ with increasing $z^+$ is evident in the low $T^+$ range ($\lesssim$ $10^4$) in figure \ref{fig1} for all three data sets, confirming that inadequate statistical convergence doesn't affect the qualitative trends exhibited by LES and SLTEST data.
In the highest possible $T^+$ range for the LES and HRNBLWT data sets, it can be noted that $|H_{L}|$ changes very slowly in magnitude for $z^+$ close to the lower-bound of the log region, and then relatively rapidly as $z^+$ nears the upper bound.
Similar behaviour is exhibited by $|H_{L}|$ from the SLTEST data, $z^+$-range for which is predominantly close to the lower-bound of the log region ($z^+_l$ $\sim$ 2600), thereby explaining its insignificant variation in the large $T^+$ range.
We can thus conclude based on figure \ref{fig1} that the qualitative trends exhibited by the LES and SLTEST data sets are physical and not an artefact of inadequate statistical convergence.
Interested readers may refer to \citet{deshpande2021} for a further detailed discussion on $H_{L}$ and the full derivation of (\ref{eq13}).

\begin{figure*}
	\centering
	\includegraphics[width=1.05\textwidth]{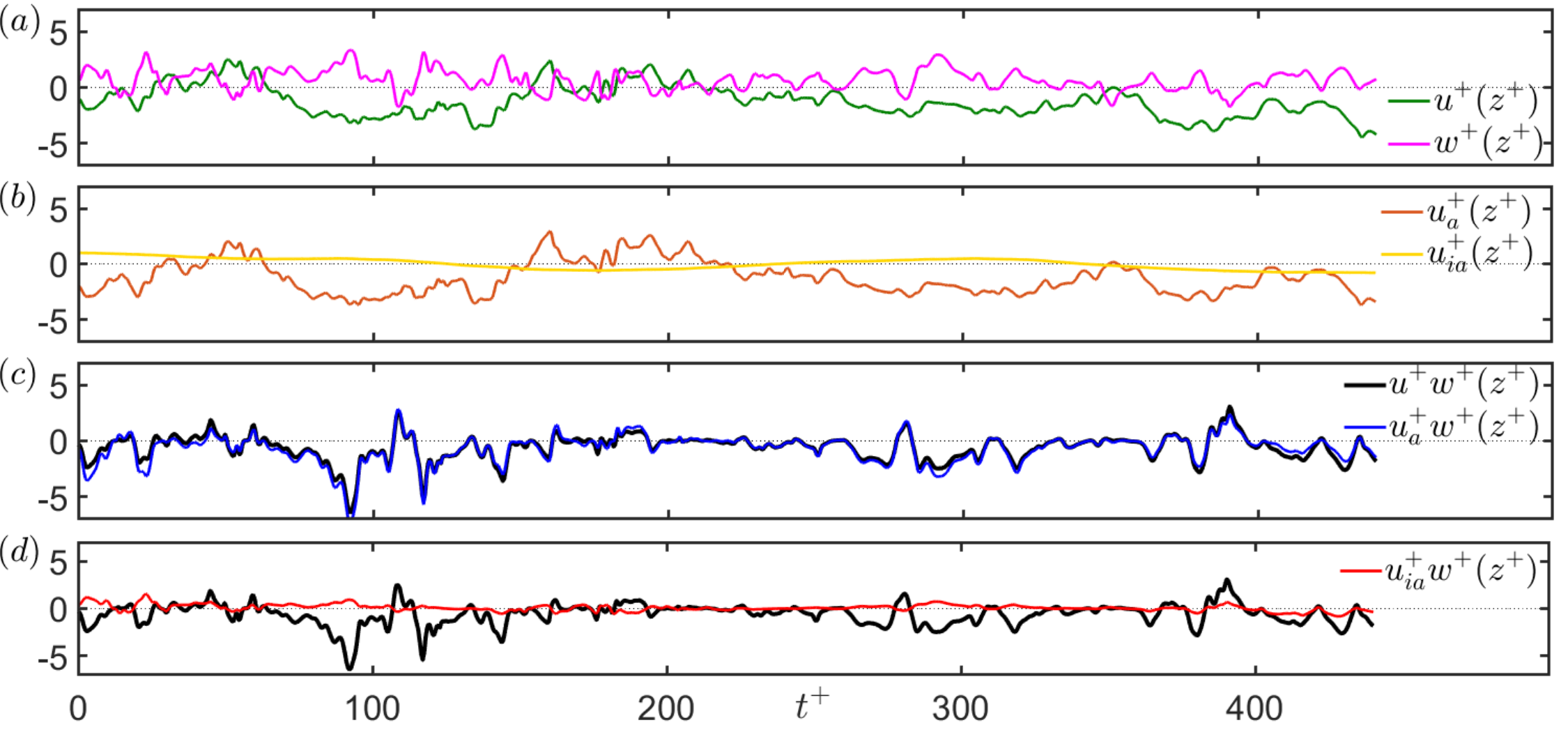}
	\caption{(a) Selected portion of synchronously acquired $u$- and $w$-time series by a cross-wire at $z^+$ $\approx$ 2.6$\sqrt{Re_{\tau}}$ from the HRNBLWT dataset.
	(b) $u$-signal decomposed into its $u_{a}$ and $u_{ia}$ components. 
(c,d) Comparison of the full Reynolds shear stress ($uw$) signal computed from $u$- and $w$-time series in (a), with its corresponding (c) ${u_{a}}{w}$ and (d) ${u_{ia}}{w}$ components.}
	\label{fig2}
	\end{figure*}

With $u_{\rm ia}$($z$) known via (\ref{eq13}), the time series of the active component ($u_a$) at $z$ can be obtained following:
\begin{equation}
\label{eq15}
\begin{aligned}
{{{u}_{a}}(z^{+};{t^{+}})} &= {{{u}}(z^{+};{t^{+}})} - {{{u}_{ia}}(z^{+};{t^{+}})},\; \text{where}\\
{{{u}_{ia}}(z^{+};{t^{+}})} &= {\mathcal{F}^{-1}}({{{u}_{ia}}(z^{+};{f^{+}})}).
\end{aligned}
\end{equation}
Figure \ref{fig2} demonstrates an example of implementing equations (\ref{eq13}), (\ref{eq14}) and (\ref{eq15}) to obtain $u_{\rm ia}$ and $u_{\rm a}$ at the log region lower-bound ($z^{+}_l$ $\approx$ 2.6$\sqrt{Re_{\tau}}$) for the HRNBLWT dataset.
This plot has been adapted from \citet{deshpande2021uw}, wherein the success of this data decomposition procedure has been demonstrated previously.
%Figures \ref{fig2}(a,b) depict a small portion of the $u$-time series at $z^{+}$ $\approx$ 2.6$\sqrt{Re_{\tau}}$ from the HRNBLWT dataset, and its corresponding active and inactive sub-components obtained using (\ref{eq13},\ref{eq15}).
In figure \ref{fig2}, $u_{\rm a}$ can be noted to resemble the small-scale (high-frequency) characteristics of $u$, which makes sense given its association with motions localized at $z^{+}$ (active).
On the other hand, $u_{\rm ia}$ represents large-scale variations owing to its association with taller motions.
This decomposition of $u$ permits comparison of the active (${u_{a}}{w}$) and inactive (${u_{ia}}{w}$) components of the Reynolds shear stresses with the total stress (${u}{w}$) in figures \ref{fig2}(c,d), respectively.
%The availability of the time series of both active and inactive streamwise components, along with the synchronously measured $w$ (figure \ref{fig1}(a)), permits the estimation of the respective Reynolds shear stress time series.
%The active (${u^{+}_{a}}{w^{+}}$) and inactive (${u^{+}_{ia}}{w^{+}}$) components of the Reynolds shear stress are respectively compared with the total (${u^{+}}{w^{+}}$) in figures \ref{fig1}(c,d).
Here, ${u_{a}}w$ can be seen to follow the $uw$-signal reasonably well, while the magnitude of ${u_{ia}}w$ is close to zero for the majority part of the signal.
This suggests that ${u_{a}}w$ is the dominant contributor to $\overline{uw}$.
The result is consistent with the original description of active motions postulated by \citet{townsend1976}, per which active motions are solely responsible for the Reynolds shear stresses.
With the characteristics of active and inactive motions established based on the well-converged spectral estimates of the HRNBLWT data set, we next proceed towards implementing this same data decomposition procedure on the LES and SLTEST data.
%, to investigate characteristics of $u_a$ and $u_{ia}$ across the log region.
	
%%%%%%%%%%%%%%%%%%%%%%%%%%%%%%%%%%%%%%%%%%%%%%%%%%%%%%%%%%%%%%%%%%%%%%
\section*{RESULTS}

\begin{figure*}
	\centering
	\includegraphics[width=1.05\textwidth]{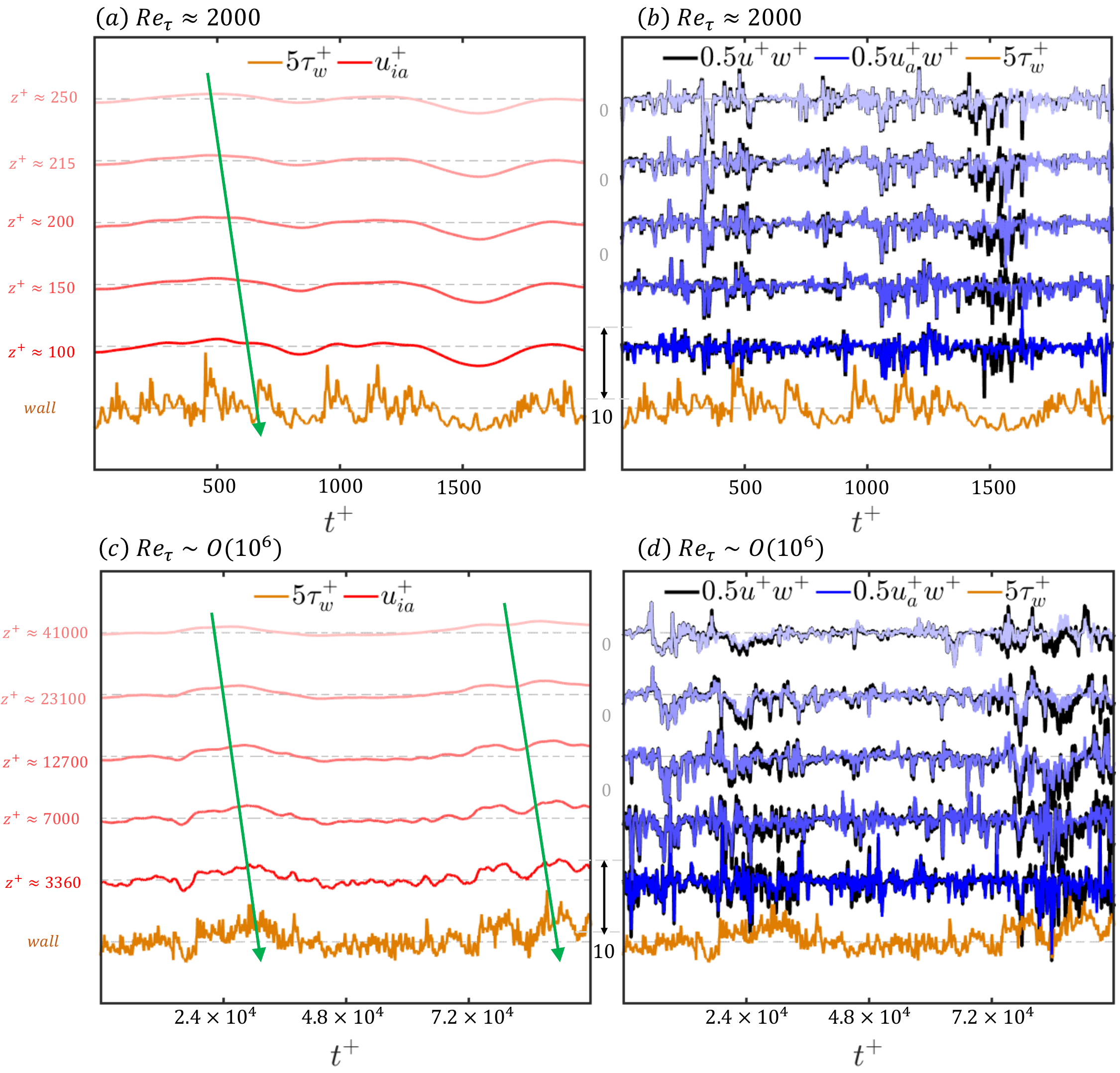}
	\caption{(a,c) Portion of the $u_{ia}$-signal estimated at various $z^+$, in the log region, from the (a) LES and (c) SLTEST datasets.
	(b,d) Portion of the ${u_{a}}w$-signal corresponding to the same time instants and $z^+$ as in (a,c), compared with the $uw$-signal at the respective $z^+$ for the (b) LES and (d) SLTEST datasets.
	Dark to light shading for these signals indicates increase in $z^+$, with each signal offset vertically as highlighted. 
	Green arrows in (a,c) are used to suggest the transfer of streamwise momentum from the log region towards the wall, with the latter represented by the ${\tau}_{w}$-signal.}
	\label{fig3}
	\end{figure*}

Decomposition of the instantaneous velocity fluctuations from LES and SLTEST data yields synchronous $u_a$ and $u_{ia}$ components across the log region, which are depicted in figure \ref{fig3}.
%In case of this dataset, notably, the log region data was acquired synchronously by sonic anemometers placed at five different wall-normal locations ($z$), thereby permitting investigation of the wall-normal variation of the instantaneous ${u_{a}}w$ and $u_{ia}$ across the log region.
%This, however, is not possible for the HRNBLWT dataset, where the \{$u,w$\} time series at each $z$ was acquired independently by traversing the cross-wire probe. 
Figures \ref{fig3}(b,d) depict ${u_{a}}{w}$-signals (in blue shading) estimated at various $z^{+}$ for the LES and SLTEST data, respectively.
These signals exhibit excellent overlap with the full Reynolds shear stress signals ($uw$; in black shading) at the corresponding $z^+$, thereby demonstrating consistency with the characteristics expected from the active motions (also exhibited previously in figure \ref{fig2}c by HRNBLWT data).  
This result reaffirms the efficacy of the SLSE-based energy decomposition scheme across a large span of $Re_{\tau}$, and instills confidence in qualitative estimates from LES and SLTEST data (based on inadequately converged $H_{L}$ at large $T^+$).

While the Reynolds shear stress (\emph{i.e.}, active) signals are de-correlated from the ${\tau}_w$-signal, the inactive signals are correlated with the ${\tau}_w$-signal by definition (refer equation (\ref{eq13})).
However, the novelty in the present analysis is revelation of the trend exhibited by synchronous $u_{ia}$ signals across the log region.
Interestingly, both the positive and negative fluctuations in the ${u_{ia}}$-signal can be seen growing in magnitude as $z^{+}$ decreases, \emph{i.e.} with reduction in distance from the wall (indicated by an arrow).
This can be associated with the increasing contributions from the attached eddy hierarchy (specifically, the inactive motions) expected with reduction in $z^+$.
Remarkably, the ${\tau}_{w}$-signal can be observed to be amplified close to those instants when $u_{ia}$ $>$ 0 in the log region.
The entire analysis in figure \ref{fig3}, thus, can be deemed as a strong empirical evidence to the hypothesis of \citet{giovanetti2016}, per which: ``\emph{the energy-containing motions, which essentially reside in the logarithmic and outer regions, transport the streamwise momentum to the near-wall region through their inactive part, while generating Reynolds shear stress with their wall-detached wall-normal velocity component in the region much further from the wall}.''
Here, the `wall-detached' part essentially refers to the active component which, as mentioned previously, is de-correlated from the wall-shear-stress signal.

%%%%%%%%%%%%%%%%%%%%%%%%%%%%%%%%%%%%%%%%%%%%%%%%%%%%%%%%%%%%%%%%%%%%%%
\section*{SUMMARY AND CONCLUDING REMARKS}
The present study implements an energy-decomposition methodology onto canonical TBL data sets, spanning a large range of $Re_{\tau}$ ($10^3$--$10^6$), to dissect the log-region flow into its active and inactive components.
Consistent results are noted across data acquired from simulations, laboratory as well as atmospheric experiments, reaffirming the physical interpretations based on the data.
Decomposition of the instantaneous velocity fluctuations unravels the unique roles played by the active and inactive components towards skin-friction drag generation, providing strong empirical evidence in support of the hypothesis of \citet{giovanetti2016}.
The present analysis reveals that the active motions are responsible for carrying majority of the Reynolds shear stresses (and subsequently, production of the turbulent kinetic energy) in the log region of a canonical TBL.
While the inactive motions are responsible for the wall-ward transport of streamwise momentum from the log region to the wall.

The present work suggests an increased significance of the `sweep' motions, which correspond to the fourth quadrant of the Reynolds shear stresses: $u$ $>$ 0, $w$ $<$ 0, in the log region of high $Re_{\tau}$ TBLs.
Interestingly, recent explainable deep learning models trained on canonical TBL data sets \citep{cremades2024} have identified these sweeps amongst the most dynamically significant coherent motions in high $Re_{\tau}$ wall turbulence.
While the growing statistical significance of sweeps has previously been quantified in high-$Re_{\tau}$ experiments \citep{deshpande2021uw}, more detailed work is currently in progress to demonstrate their important role in the energy transfer mechanisms of the outer region of TBLs.
Another interesting line of future research could be towards investigating the relationship (if any) between Townsend's active and inactive motions, with the dynamically `important' coherent structures identified by explainable deep learning models \citep{cremades2024}.
Besides the aforementioned developments associated with our fundamental understanding, the present work also carries the potential to inform the design of future energy-efficient flow-control strategies.

\section*{ACKOWLEDGEMENTS}
R. D. is supported by University of Melbourne's Postdoctoral Fellowship. Funding is gratefully acknowledged from ONR: N62909-23-1-2068 (R.D., I.M.), and to R.V. from ERC grant no. `2021-CoG-101043998, DEEPCONTROL'.

\bibliographystyle{tsfp}
\bibliography{Rahul_tsfp2024_FullPaper_ArXiv}

%%%%%%%%%%%%%%%%%%%%%%%%%%%%%%%%%%%%%%%%%%%%%%%%%%%%%%%%%%%%%%%%%%%%%%
%%%% FIGURES %%%%%
%\newpage

\end{document}